\documentclass[a4paper]{jpconf}
\usepackage{graphicx}
\begin{document}
\title{The DMTPC project  }

\author{Gabriella Sciolla}
\address{ Massachusetts Institute of Technology, 
 77 Massachusetts Ave., Cambridge, MA 02139, USA}
\ead{sciolla@mit.edu}

\author{for the DMTPC Collaboration}
\address{ J.~Battat, T.~Caldwell, D.~Dujmic, P.~Fisher, S.~Henderson, R.~Lanza,  A.~Lee, 
J.~Lopez, A.~Kaboth, G.~Kohse,  J.~Monroe, T.~Sahin, G.~Sciolla, R.~Yamamoto,  
H.~Yegoryan (Massachusetts Institute of Technology);  
S.~Ahlen, K.~Otis, H.~Tomita (Boston University); 
A.~Dushkin, H.~Wellenstein (Brandeis University). }
\begin{abstract}
The DMTPC detector is  a low-pressure CF$_4$ TPC with optical readout for 
directional detection of  Dark Matter.  
The combination of the energy and directional tracking information allows for 
an efficient suppression of all backgrounds. 
The choice of gas (CF$_4$) makes this detector particularly sensitive 
to spin-dependent interactions. 
\end{abstract}

\section{Directional Dark Matter detection}

The goal of a directional Dark Matter detector~\cite{sciolla2008} 
is to provide an unambiguous observation of Dark Matter (DM) 
by measuring both the direction and the energy of the nuclear recoil 
produced by the interaction of a  Weakly Interacting Massive Particle (WIMP) 
 with a nucleus in the active mass of the detector.

The simplest models of the distribution of WIMPs in our Galaxy suggest that the orbital motion of the Sun about 
the Galactic center will result in  an 
Earth-bound observer to experience a WIMP wind with speed 220 km/s (the galacto-centric velocity of the Sun) 
originating from the direction of the Sun's motion.
Because the Earth's rotation axis is 
oriented at 48$^\circ$  with respect to the direction of the WIMP wind,
the average  direction of DM particles recorded by an Earth-bound observer changes 
by about 90$^o$ every 12 hours~\cite{spergel}. 
The ability to measure such a direction would provide a powerful suppression of insidious backgrounds 
such as cosmogenic neutrons and neutrinos from the Sun~\cite{monroe}, 
as well as a unique instrument to test local DM halo models~\cite{Alenazi,MorganGreenSpooner}.
This capability  makes directional detectors unique observatories for underground WIMP astronomy. 

\begin{figure}[t]
\centering
\includegraphics[width=4in]{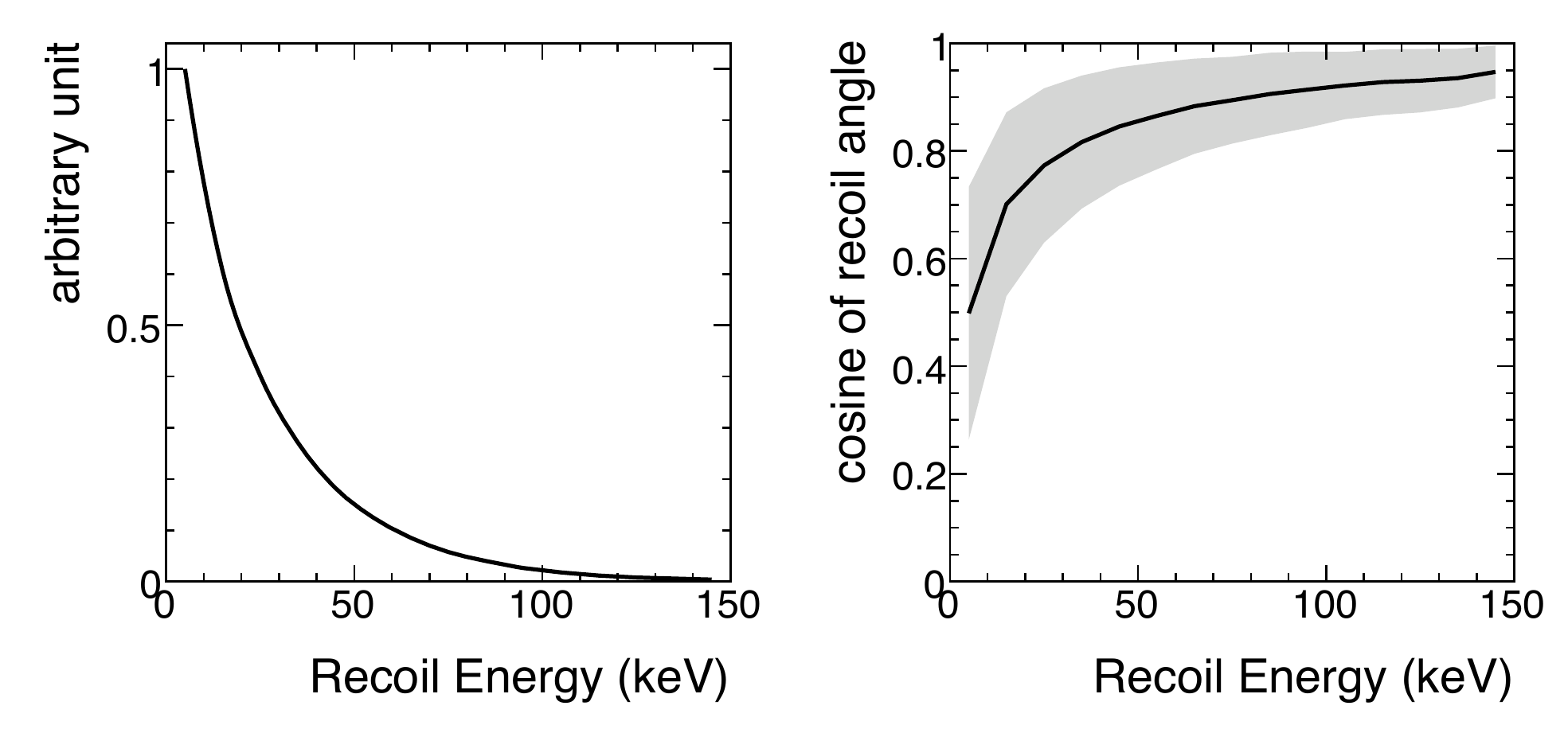}
\vspace{-10pt}
\caption{
Energy spectrum and distribution of the cosine of the recoil angle
  vs. recoil energy for fluorine  recoils induced by 100~GeV WIMPs in CF$_4$. 
The recoil angle is defined as the angle between the direction of the 
recoil and the direction of the incoming particle. 
} \label{Fig1}
\end{figure}


 
 Dark Matter interactions in the detector   generate nuclear recoils with typical 
energies of  few tens of keV 
(Fig.~\ref{Fig1}, left). The direction of the recoiling nucleus encodes the direction of the incoming DM particle 
(Fig.~\ref{Fig1}, right). To  observe the daily modulation in the direction of the DM wind, 
an  angular resolution  of  $\approx30^\circ$ in the reconstruction of the recoil nucleus is required. 

Directional DM experiments use low-pressure (40--100 torr) gas as target material, in which typical DM induced 
nuclear recoils have a length of a few millimeters.  
A 3-D reconstruction of the recoil track with a spatial resolution of several hundred microns in all three coordinates 
is sufficient to achieve the desired angular resolution.   
A 2-D reconstruction is still valuable, although it slightly degrades the sensitivity~\cite{GreenMorgan}. 
``Head-tail'' discrimination  
is also very important as it  improves 
the sensitivity for detecting DM by about one order of magnitude~\cite{GreenMorgan}. 

Due to the low-density of the target material, directional detectors tend to be large in volume: 
a typical directional detector with a fiducial mass of one ton occupies O(10$^3$) cubic meters. 
Therefore,  the success of a directional DM program depends very strongly on the developement of  detectors 
with a low cost per unit volume. 
The largest expense for standard gaseous detectors is the cost of the readout electronics,
 making it essential to utilize low-cost readouts in order to make DM directional detectors financially viable.

\section{The DMTPC detector concept }   

\begin{figure}[tb]
\centering
\includegraphics[width=3in]{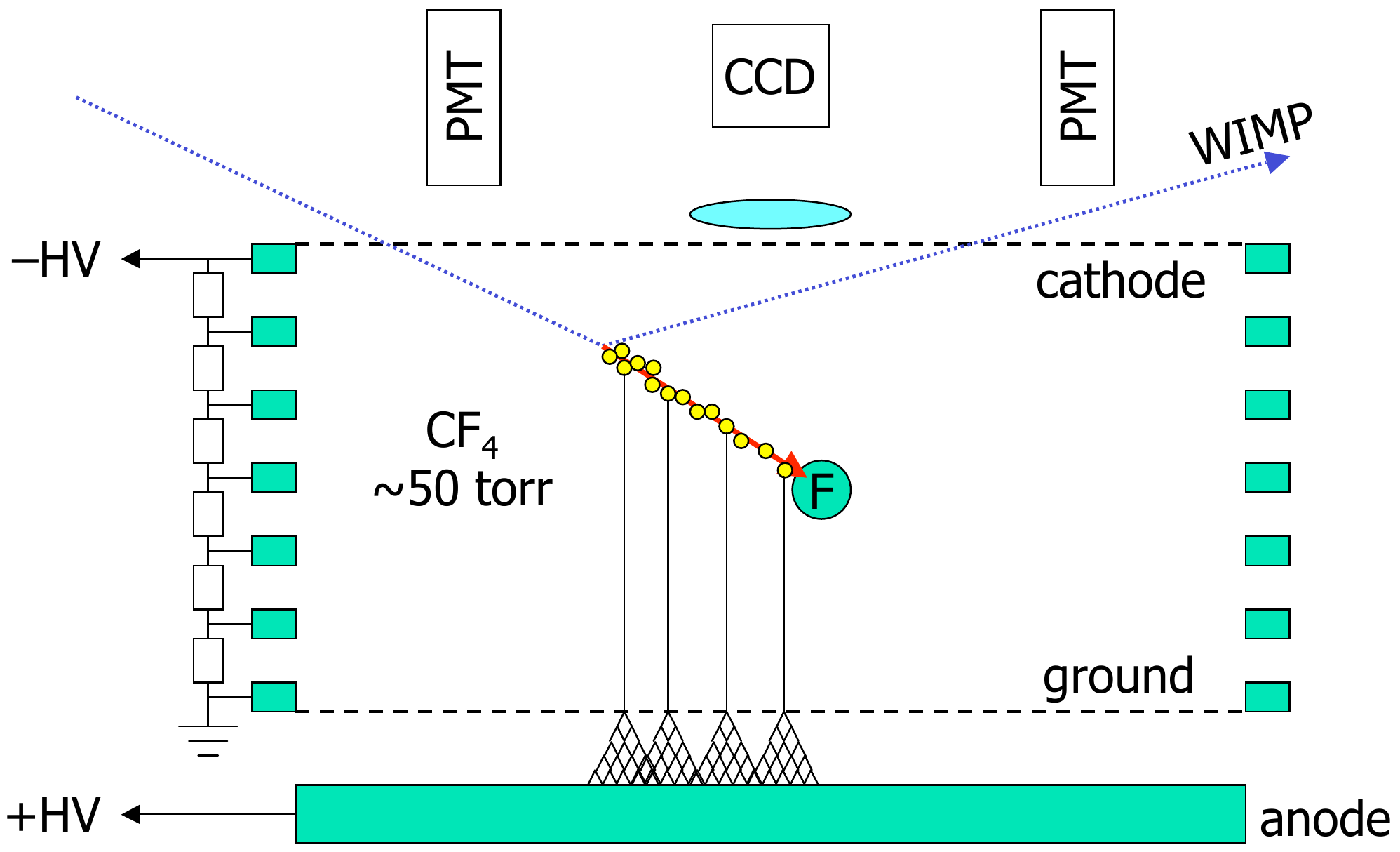}\qquad
\raisebox{0.2in}{\includegraphics[width=3in]{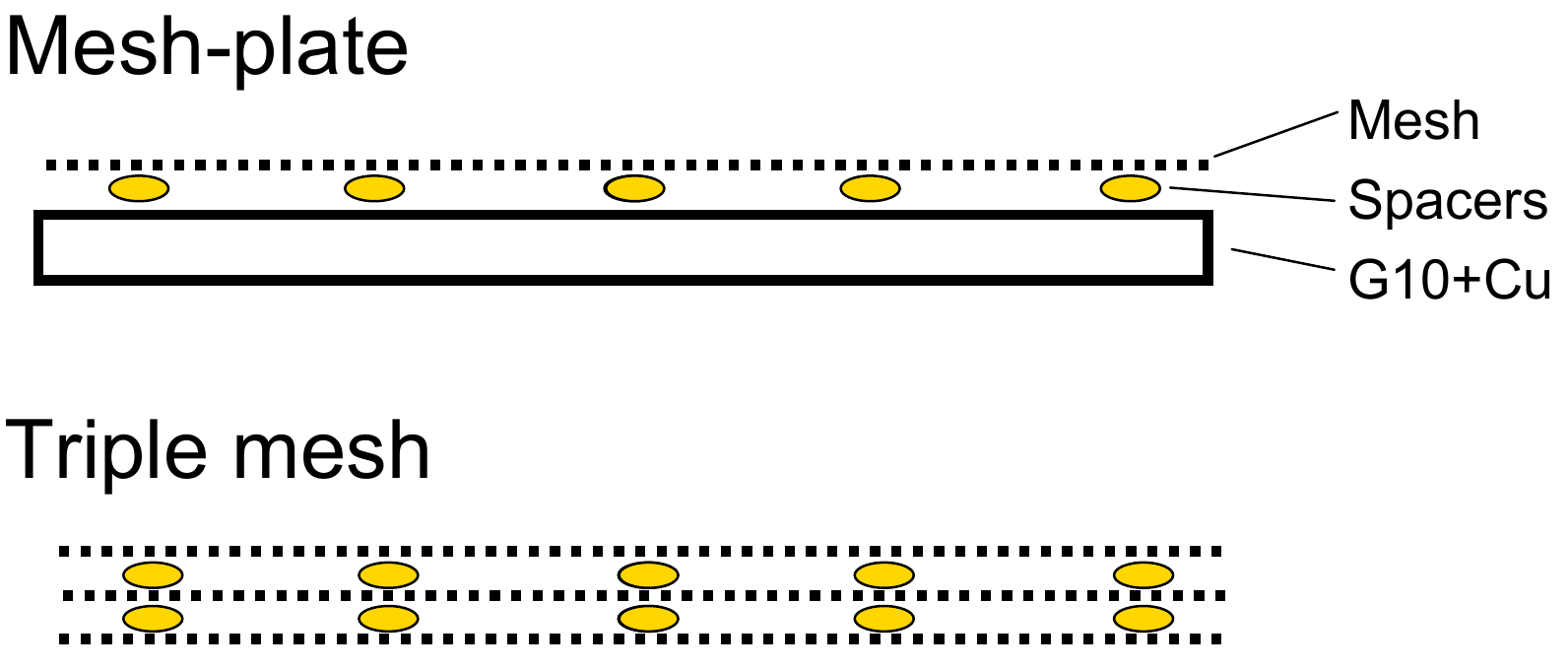}}
\vspace{-10pt}
\caption{ 
Left: illustration of the DMTPC detector concept. 
Right: two designs for the amplification region: ``mesh-plate''  (top) and ``triple mesh'' (bottom). 
} \label{amplification}
\end{figure}

The DMTPC detector is a low-pressure TPC with optical readout. 
The detection principle is illustrated in Fig.~\ref{amplification}, left. 
The TPC is filled with 
CF$_4$ at a pressure of about 50 torr. 
At this pressure, a typical 50 keV nuclear recoil has a length of about 2 mm. 
The average ionization energy in CF$_4$ is 54 eV~\cite{NEWAGE2},  which results in 
about $10^3$  primary electrons from the nuclear recoil. 
Such electrons 
drift toward the amplification region, where 
photons are produced together with electrons in the avalanche process. 
A CCD camera mounted above the cathode mesh records the image of 
the  nuclear recoil projected along the amplification plane. 
The sense of the recoiling nucleus 
can be determined   (``head-tail'' discrimination), because the energy loss (dE/dx) is not constant along the trajectory. 
An array of photomultipliers (PMTs) mounted above the cathode mesh 
determines the length of the recoil in the drift direction through time measurements.

 CF$_4$ was chosen as  the target material primarily because of its high fluorine content. Fluorine is an
excellent element to detect spin-dependent interactions on protons~\cite{ellis}, 
due to its large spin factor and isotopic abundance 
 (Table~\ref{tableF}). 
Spin-dependent (SD) interactions are predicted to dominate over spin-independent ones in  models
where the lightest super-symmetric particle has a substantial Higgsino contribution~\cite{SDTheory,ellisSD}. 


CF$_4$ is also an excellent detector material~\cite{Pansky,Christophorou}. Its 
good scintillation properties are very important for the  
optical readout. Our recent measurements~\cite{asher} indicate that in CF$_4$  
the number of photons produced 
between 200 and 800 nm wavelength
is about 1 for every 3 avalanche electrons. 
Moreover, the low transverse diffusion characteristic of electrons in CF$_4$ allows for  good spatial 
resolution in the reconstruction of the recoil track despite the long (25 cm) electron drift distance. 
Finally, CF$_4$ is non-flammable and non-toxic allowing for  safe operation underground.

%
%

\begin{center}
\begin{table}[b]
\centering
\caption{\label{jfonts} Spin of the nucleus J, nuclear spin factor $\lambda^2$J(J+1), and 
  abundance in nature of various isotopes considered for 
  SD-interaction studies. The figure of merit 
 is defined as the product of the square of the nuclear mass, 
  the number of isotopes per kg, and the  spin factor.}
\label{tableF}\medskip
\begin{tabular}{@{}l*{15}{l}}
\br
  Isotope  &  J    &  $\lambda^2$J(J+1)  & Natural abundance & Figure of merit\\
\mr
 $^{19}$F  & 1/2  &  	0.65 &  100\% & 	74     \\
 $^{23}$Na & 3/2  &  	0.04 &  100\% & 	6      \\
 $^{73}$Ge & 9/2  &  	0.06 &  7.8\% & 	2      \\
 $^{93}$Nb & 9/2  &  	0.16 &  100\% & 	91     \\
 $^{127}$I & 5/2  &  	0.01 &  100\% & 	5      \\
 $^{129}$Xe & 1/2  &  	0.12 &  26\% & 	25             \\
\br
\end{tabular}
\end{table}
\end{center}

CCDs provide a true 2-D readout at
a much lower cost-per-channel than any other readout technology used in particle physics. 
A modern low-noise CCD camera with 4 megapixels can be purchased today for a few thousand US dollars, 
which corresponds to $10^{-3}$ dollars/channel. The cost of a directional DM detector is dominated by 
the readout electronics, making the optical readout a solution toward an economically viable detector.

Two alternative implementations of the amplification region~\cite{meshpaper}  are 
illustrated in Fig.~\ref{amplification}.  
In the ``mesh-plate'' design (top) the amplification is obtained by 
applying  a large potential difference 
($\Delta$V = 0.6--1.1 kV) 
between a copper plate and a conductive woven mesh. A uniform distance 
of 0.5 mm between the plate and the mesh 
is ensured by the use of fishing wires spaced 2 cm apart. 
The copper or stainless steel mesh is made of 28 $\mu$m wire with a pitch of 256 $\mu$m. 
The pitch of the mesh determines the intrinsic spatial resolution of the detector. 
The ``triple mesh'' design (bottom) 
yields a transparent amplification region, 
which enables a single CCD camera to image two drift regions, leading to a substantial cost reduction.


The DMTPC detector measures 
 the number of photons in the CCD camera, 
 the 2-D image of the recoiling nucleus,  
 the distribution of energy loss along the recoil track,  
 the width and integral of the PMT signal, and 
 the electronic signal produced on the amplification plane.   

The energy of the nuclear recoil is independently determined  
by  the measurement of the number of photons observed in the CCDs, 
the integral of the electronic signal produced on the amplification mesh, 
and the integral of the PMT signal. The redundancy in the design is intentional 
and has the goal of maximizing the robustness of the measurement. 
The track length and direction of the recoiling nucleus are reconstructed 
by combining the measurement of the projection along the amplification 
plane (from pattern recognition in the CCD) and the projection along the 
direction of drift (from the width of the signal recorded in the PMTs). 
The sense of the recoil track can be determined by the Bragg curve. 
The PMTs will also provide a measurement of the absolute z position of the recoil by measuring the 
time between  the primary ionization and the amplified signal. This information will 
drastically reduce possible backgrounds from the amplification meshes. 

The CCD images have a long (1--10 seconds) exposure. 
If during the exposure a trigger is generated by the PMT or the 
electronic readout of the amplification plane, the CCD is read out and the event is saved to disk. 
Otherwise, the CCD is reset, minimizing dead time. 

The combination of the measurements described above is very effective in reconstructing the energy, direction, 
and sense of  nuclear recoils from WIMPs. 
In addition,  an excellent rejection of the $e^-$ and $\alpha$ backgrounds can be obtained 
by combining the measurement of the energy and length of the recoil (see below).

\section {R\&D results: demonstration of directional detection }

%

\begin{figure}[b]
\centering
\includegraphics[width=0.8\textwidth]{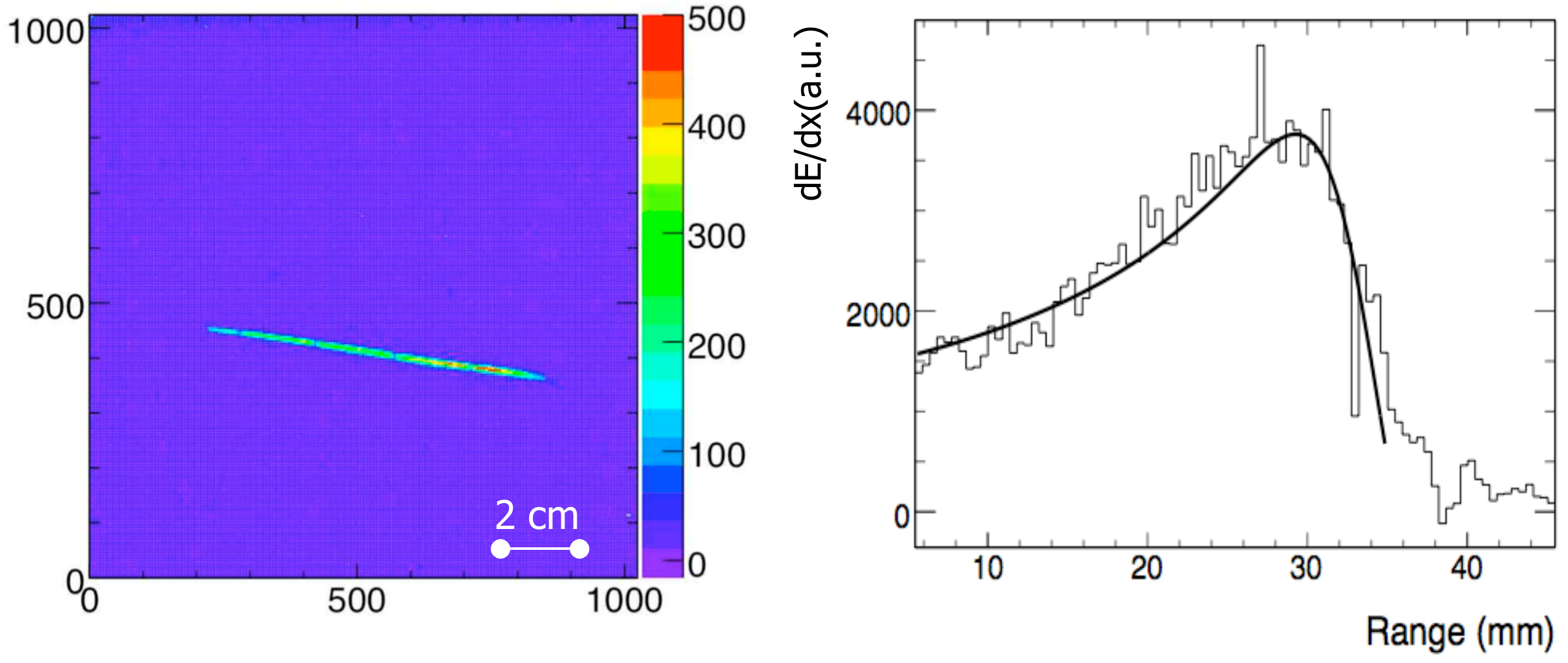}
\vspace{-10pt}
\caption[]{
Left: an $\alpha$ track stopped in  CF$_4$ at  100~torr (direction: left to right). 
Right: average light yield  vs. range of the $\alpha$ tracks in data (histogram) 
compared to the SRIM MC prediction~\cite{SRIM}. } \label{alpha}
\end{figure}

An $^{241}$Am source that produces 5.5 MeV alpha particles is used 
as calibration source. 
The image of one alpha particle stopped inside CF$_4$ at 100~torr is shown in Fig.~\ref{alpha}, left. 
Fig.~\ref{alpha}, right, shows the Bragg curve for 5.5 MeV alpha particle in CF$_4$ at 280 torr. 
The histogram shows the  energy loss measured in the  prototype, while the solid line shows 
the results of a SRIM~\cite{SRIM} simulation. The agreement between data and MC is better than 10\%. 
The alpha source is used to study the gain of the detector as a function of the voltage 
in the amplification region and gas pressure, while a $^{55}$Fe can be used for absolute 
energy calibration. 
The alpha source is also used to measure the resolution as a function of the drift distance of the 
primary electrons to quantify the effect of the diffusion. Our studies~\cite{headtailpaper} 
show that the transverse diffusion has $\sigma<$ 1 mm for a drift distance of 25 cm. 

\begin{figure}[t]
\centering
\includegraphics[width=0.9\textwidth]{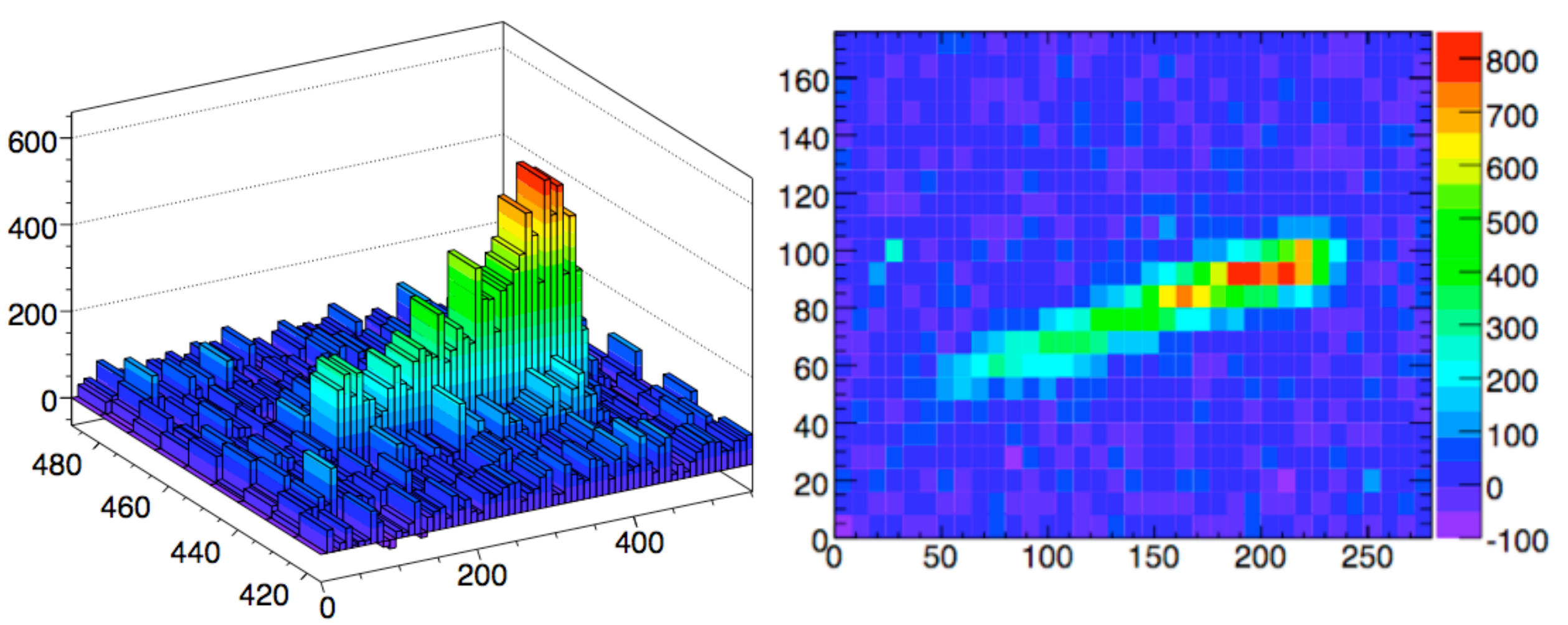}
\vspace{-10pt}
\caption[]{
Left: nuclear recoil from a 14 MeV neutron in CF$_4$ at 250~torr recorded with the wire-based DMTPC prototype. 
Right: nuclear recoil induced by a neutron from a $^{252}$Cf source in  CF$_4$ at 75~torr. 
In both cases the neutrons were traveling right to left. 
The higher $dE/dx$ visible on the right of the track is consistent with observation of the ``head-tail'' effect. 
} \label{recoils2d}
\end{figure}

The first demonstration of the DMTPC detector concept was obtained 
with a small chamber that used parallel wire planes to obtain charge amplification. 
This detector was used to  perform the first observation of 
the ``head-tail'' effect in nuclear recoils generated by low-energy neutrons~\cite{headtailpaper}.  
The determination of the sense and direction of nuclear recoils was evaluated by studying the recoil 
of fluorine nuclei in interactions with low-energy neutrons. 
For nuclear recoils  below 1 MeV only the tail of the Bragg peak is visible, and therefore 
the energy deposition decreases along the path of the recoil, allowing for the 
identification of  the ``head'' (``tail'') of the event by a smaller (larger) energy deposition. 
This initial measurement used 14 MeV neutrons from a D-T generator. 
The reconstructed recoils had energy between 200 and 800 keV. 
The energy distribution along the recoil track for one of such recoils is shown in Fig.~\ref{recoils2d}, left. 
We measured that $(73\pm3)$\% of the recoils had the correct sense. 
This measurement represents an observation of the  
the ``head-tail'' effect  with a significance of  8 $\sigma$~\cite{headtailpaper}. 

\begin{figure}[b]
\center
\includegraphics[width=0.45\textwidth]{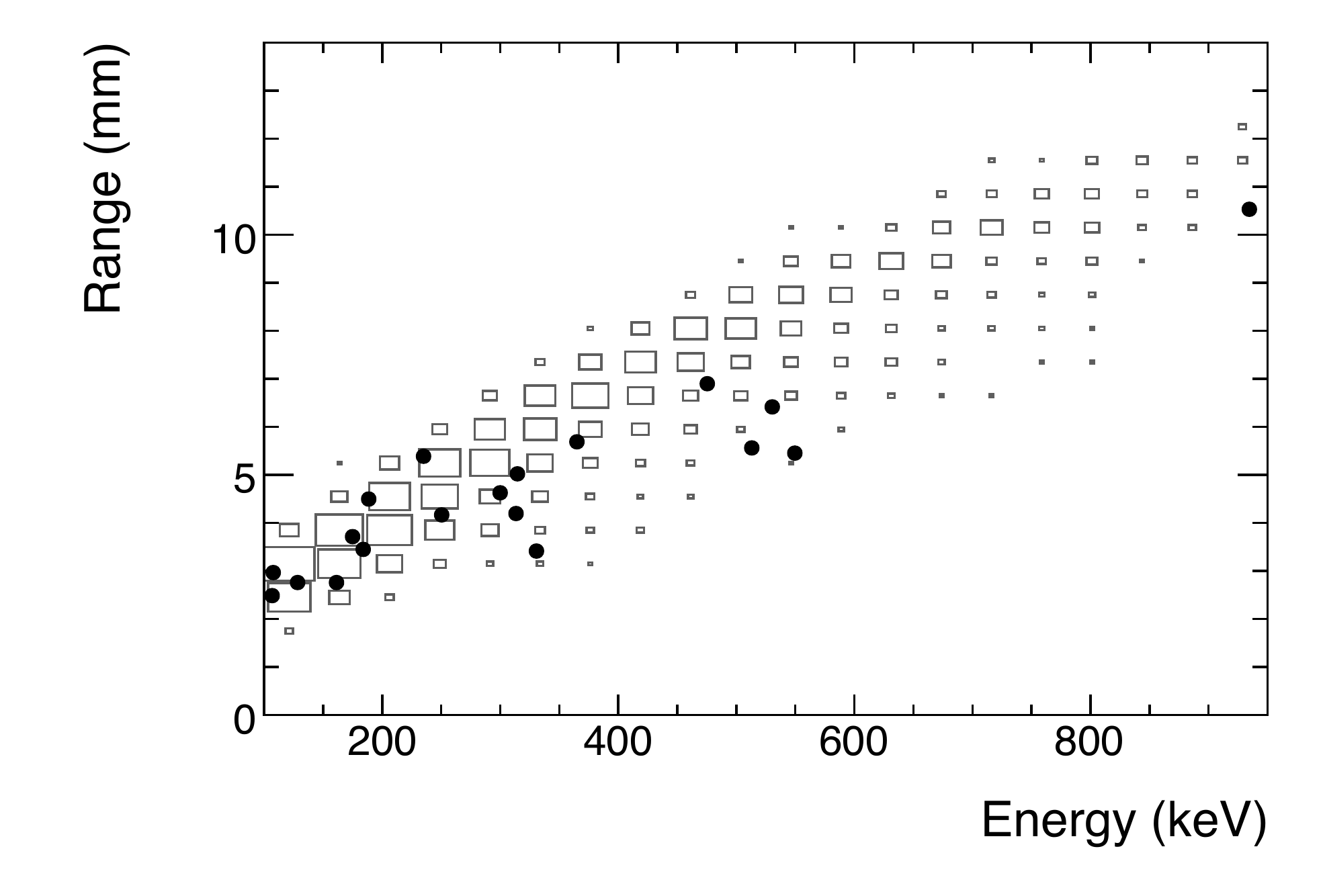}%
\includegraphics[width=0.45\textwidth]{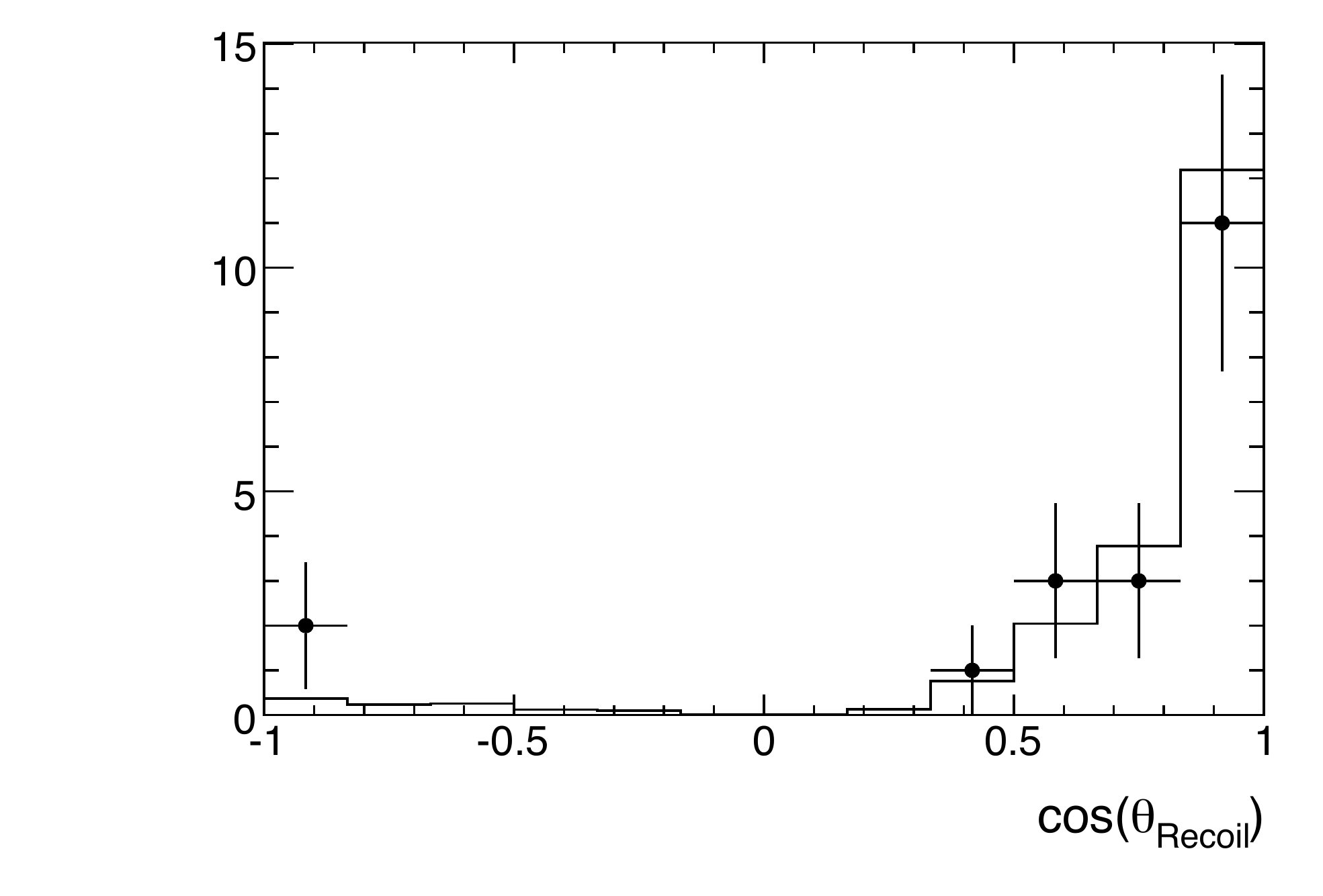}%
\vspace{-10pt}
\caption{Range  
vs. reconstructed energy (left) 
and signed cosine of the recoil angle (right) of  nuclear recoil candidates from $^{252}$Cf  at 75~torr. 
Black points are data; histogram is simulation.
\label{CfRun}}
\end{figure}

A newer mesh-based detector was later used to study the ``head-tail'' effect in 
lower energy neutrons generated by a $^{252}$Cf  source. 
This detector allowed us to obtain a  2-D reconstruction of the nuclear recoils.   
Better sensitivity  to lower energy thresholds was achieved by lowering the CF$_4$ pressure to 75 torr. 
Fig.~\ref{recoils2d}, right shows a Cf-induced nuclear recoil 
due to a neutron traveling right to left inside our detector. 
 The decreasing $dE/dx$ along the track direction clearly visible in the image  
proves that the detector is able to determine the sense of the direction on an event-by-event basis. 
Measurements of the track length  as a function of the recoil energy
and of the recoil angle are shown in Fig.~\ref{CfRun}. The data (black dots) 
are in good agreement with the predictions of the SRIM~\cite{SRIM} MC (histogram). 
Our recent measurements~\cite{meshpaper} demonstrated ``head-tail'' discrimination down to 100 keV at 75 torr. 
Monte Carlo studies indicate that this effect can be observed for 
recoils above 50 keV at a pressure of 50~torr when using Apogee U6 CCD cameras.  

The current DMTPC prototype 
consists of two optically independent regions contained in one stainless steel vessel. Each region is a cylinder 
with 25 cm diameter and 25 cm height. 
A field cage made of stainless steel rings keeps the uniformity of the electric field within 1\% in the 
fiducial volume. 
The amplification is obtained by using a ``mesh-plate'' design (Fig.~\ref{amplification}, top-right). 
The detector is read out by two CCD cameras, 
each imaging one drift region. The optical system consists of two Nikon photographic lenses with f-number of 1.2 
and focal length of 55 mm, 
and two Apogee U6 CCD cameras~\cite{apogee} equipped with  Kodak 1001E CCD chips. 
Because the total area imaged is $16\times16$~cm$^2$, the detector has an active volume of about 10 liters.

\section{DMTPCino detector design} 

\begin{figure}[b]
\centering
\includegraphics[width=120mm]{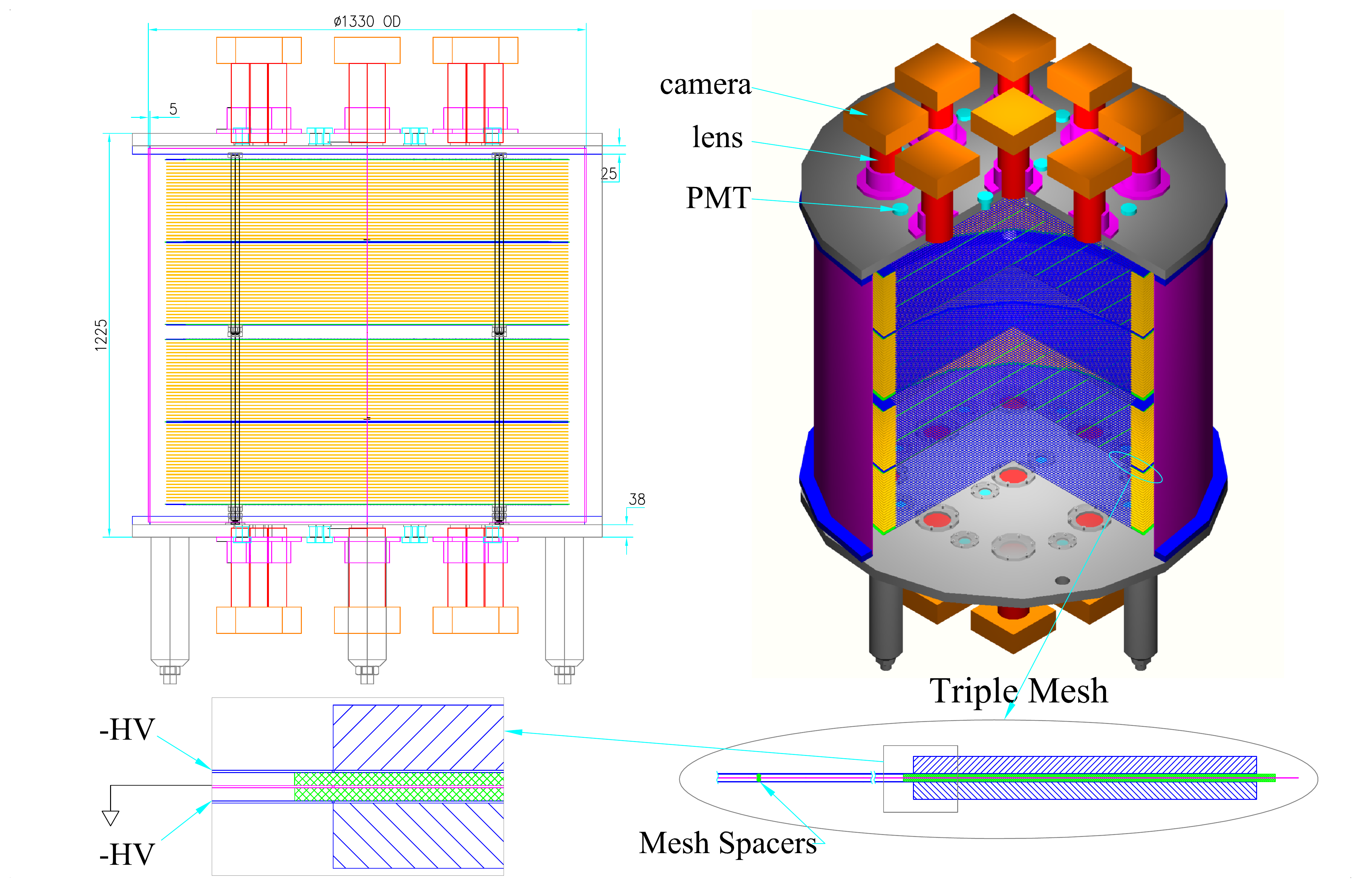}
\vspace{-10pt}
\caption{Preliminary drawings of the 1 $m^3$ DMTPC detector. } 
\label{DMTPC1m}
\end{figure}

DMTPCino is a 1-m$^3$ DMTPC detector designed to substantially improve 
limits on spin-dependent interactions of WIMPs on protons while 
testing the DMTPC detector concept on a larger scale and in a realistic environment. 
The preliminary design of the detector is shown in Fig.~\ref{DMTPC1m}. 
The apparatus consists of a stainless steel vessel of 1.3 m diameter and 1.2 m height.
Nine CCD cameras and nine PMTs are mounted on each of the top and bottom plates of the vessel, 
separated from the active volume of the detector by an acrylic window.  The detector consists of 
two optically separated regions. Each region is equipped with a triple-mesh amplification 
device, mounted in between two symmetric drift regions.  Each drift region has a diameter of 1.2 m 
and a height of 25 cm, for a total active volume of 1\,m$^3$. 
A field cage 
keeps the uniformity of the electric field 
within 1\% in the fiducial volume. A gas system circulates and purifies the CF$_4$. 

All materials used inside the active volume of the detectors are selected  to be radio-pure 
to limit backgrounds from internal radioactivity. 
Pure copper, stainless steel, Ph bronze, Vesper, quartz and acrylic  
are known to satisfy these requirements~\cite{exopaper}.  
Because all CCD cameras, lenses, and PMTs are outside the active volume, 
their contribution to  internal radioactivity is negligible. 

At a pressure of 50 torr and 21 degrees C, this module will contain 250 g of CF$_4$. 
Assuming an overall data-taking efficiency of 50\%, a one-year  run will yield an exposure of 
45 kg-days.

The CCD cameras we plan to use for DMTPCino are the same Apogee Alta U6 that we are currently using 
on the 10-liter detector. The U6 uses a Kodak Grade 2 KAF-1001E (1024$\times$1024 pixel array, 
each pixel 24$\times$24 $\mu$m$^2$) full-frame sensor with peak  quantum efficiency of 72\% at 560 nm, 
and a  readout noise of 13 (20 max) electrons.

\section{Background suppression}

The main source of background for a DMTPC detector, as for any DM experiment,  
is due to electromagnetically interacting particles such as photons and electrons, and to $\alpha$ particles. 
These particles are produced by natural radioactivity of the detector components and surrounding materials,  
as well as by cosmic rays. 
The impact of these backgrounds will be  substantially suppressed  
by using only radio-pure  materials in the fabrication of the apparatus and by  shielding  the detector.  
In addition, the directional information  provides a powerful signature 
that can be used to distinguish particle species: the correlation between the energy and length of the recoil track. 

\begin{figure}[tb]
\center
\includegraphics[width=8.5cm]{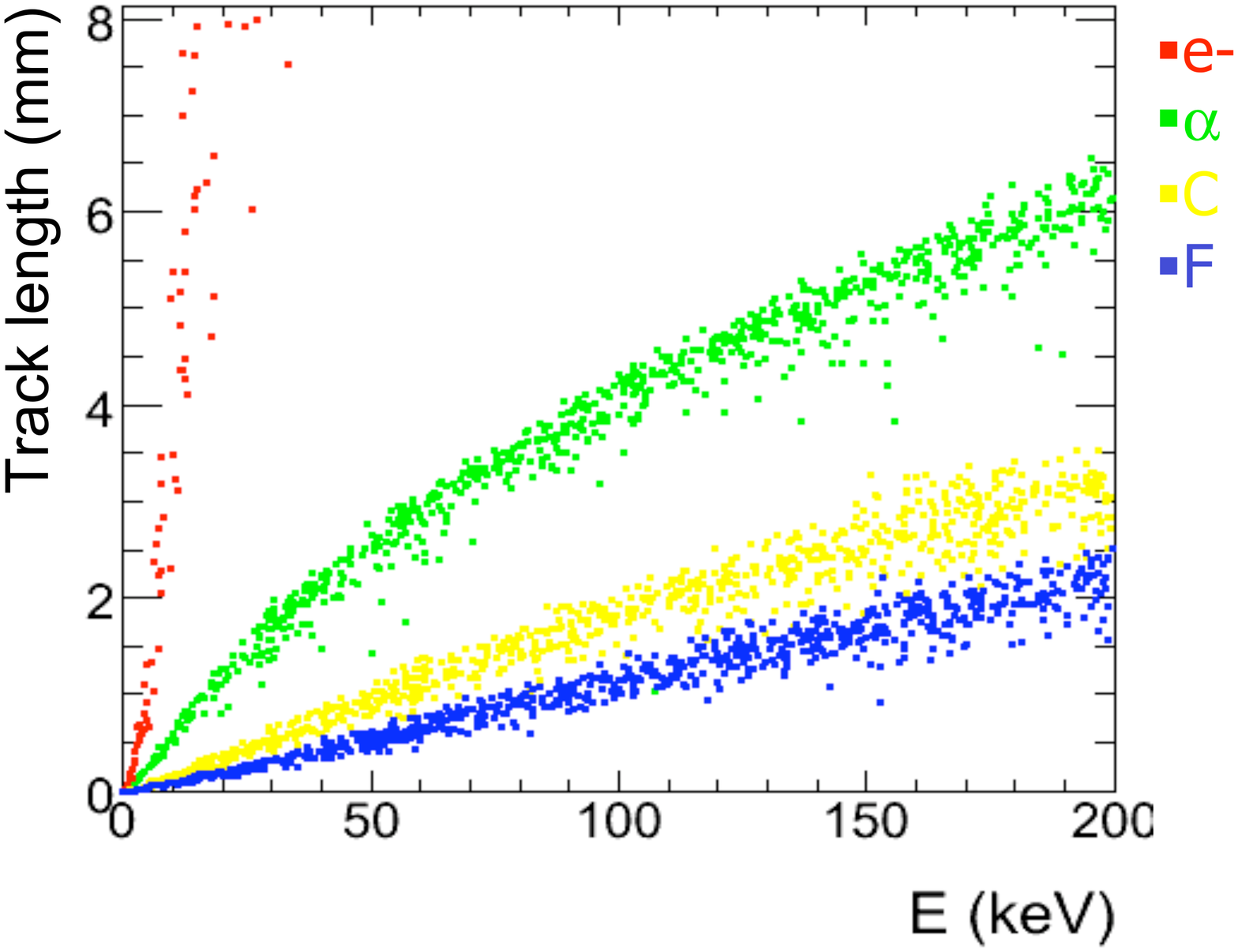}%
\includegraphics[width=7cm]{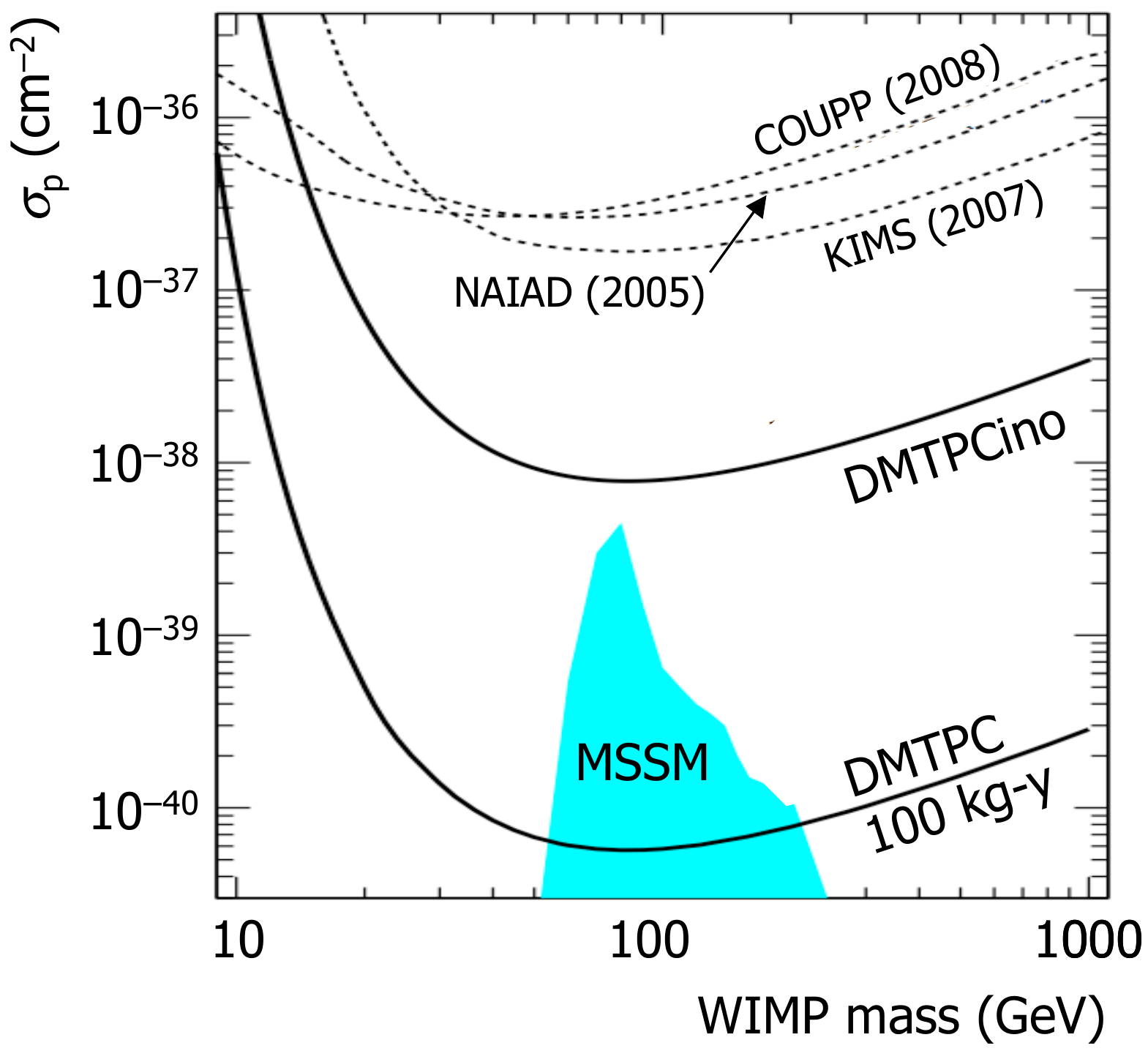}%
\vspace{-10pt}
\caption{ 
Left: range versus energy of the recoil track for  fluorine recoils (blue), carbon recoils (yellow), 
alphas (green), and electrons (red). 
Right: expected 90\% C.L. limits on  SD interactions of WIMPs 
on protons for DMTPCino (top solid line) and a larger DMTPC detector (bottom solid line). 
The dashed lines show the best present limits. 
The blue area shows the MSSM expectation.   
\label{sensitivity}}
\end{figure}

Fig.~\ref{sensitivity}, left, shows the distribution of the length versus energy 
of the track expected in the DMTPC detector 
for electrons (red), alpha particles (green), and carbon (yellow) and fluorine (blue) nuclear recoils 
from WIMP interactions. The MC simulation used is known to reproduce current data within 10\%. 
From this plot, we conclude that an excellent discrimination against $\alpha$ particles can be achieved with a 
threshold of 30 keV or lower. 
The gamma ray rejection factor, measured using a $^{137}$Cs source, is better than 2 parts per million.

Neutrons  backgrounds will be reduced by operating the detector underground and 
making use of a passive shielding. Residual neutrons will be rejected using the directional information. 

Directionality is also key to distinguish between WIMP signal and   
coherent  scattering of solar neutrinos, which  will point back to the sun.

\section{Expected sensitivity from DMTPCino}

The sensitivity of the DMTPCino detector to SD interactions of WIMPs on protons has been studied 
assuming that the detector will be operated at a threshold of 
30 keV\footnote{Assuming a 
quenching factor of about 55\% for recoils of 50 keV, as predicted in reference~\cite{hitachi}, 
a  30 keV threshold in the 
detector corresponds to a nuclear recoils with kinetic energy of 50 keV.}   
in an underground laboratory at a depth $<$ 1,600 m.w.e.,  and that 
it will be surrounded by a neutron shielding consisting of 40 cm of polyethylene. 
No significant internal backgrounds are assumed to be present above threshold. 
The limits expected from DMTPCino (5 months of data-taking at 50\% efficiency) 
are shown in  Fig.~\ref{sensitivity}, right. Improvements of a factor of 50 over the 
best existing measurements~\cite{KIMS2007, COUPP2008, NAIAD2005}  
can be obtained  despite 
the limited target mass due to the large SD cross-section of fluorine (Table~\ref{tableF}) and to the excellent 
background rejection capability of the detector. 

  
A large DMTPC detector, with an active mass of $10^2$--$10^3$ kg, will be able to explore 
a significant portion of the Minimal Supersymmetric Standard Model (MSSM) parameter space~\cite{ellisSD} 
(Fig.~\ref{sensitivity}, right). 
Such a  detector is an ideal candidate for the DUSEL laboratory. 

\section{Conclusion}
The DMTPC detector is designed to measure the 
energy, position, direction, and sense of low-energy nuclear recoils
generated by a WIMP interaction.   
The combination of the energy and tracking information allows for 
an efficient suppression  not only of backgrounds due to electrons, 
photons, and alphas, but also of more insidious backgrounds, 
such as neutrons and neutrinos. 
The choice of gas (CF$_4$) makes this detector particularly sensitive to spin-dependent interactions. 
We estimate that with a 1-m$^3$ detector (DMTPCino) in less than one year of data-taking underground 
we will be able to improve the existing  limits on SD interactions on protons by about a factor 50. 
Such a  detector will lay the foundation for the design of a large directional DM detector 
that could provide an  unambiguous positive observation of WIMPs 
as well as a way to discriminate between the various galactic DM halo models. 

\section{Acknowledgments}
This work was supported by the Advanced Detector Research Program of the U.S. Department of 
Energy, the National Science Foundation, the Pappalardo Fellowship program, 
the MIT Kavli Institute for Astrophysics and Space Research, 
and the Physics Department at the Massachusetts Institute of Technology.

\end{document}